\documentclass[twocolumn]{aastex61}

\usepackage{rotating}
\usepackage{amssymb}
\usepackage{amsfonts}
\usepackage{amsmath}
\usepackage[varg]{txfonts}
\usepackage{natbib}
\usepackage{color}
\usepackage{comment}
\usepackage{graphicx,rotating}

\usepackage{longtable}
\usepackage{threeparttablex}
\usepackage{lineno}

\newcommand{\RNum}[1]{\uppercase\expandafter{\romannumeral #1\relax}}
\newcommand{\lamost}[1]{\textsc{F\small{AMOST}}}

\newcommand\aastex{AAS\TeX}

\received{\today}
\revised{---}
\accepted{---}
\submitjournal{AAS Journal}

\shorttitle{\aastex\ Zong et al. (2018) LK-project spectra}
\shortauthors{Zong et al. (2018)}

\begin{document}

\title{LAMOST observations in the {\sl Kepler} field. \RNum{2}. Database of the low-resolution spectra {from the five-year regular survey}\footnote{Based on observations collected with the Large Sky Area Multi-Object Fiber spectroscopic Telescope (LAMOST) which is located at the Xinglong Observatory, China.}}

\correspondingauthor{Jian-Ning Fu}
\email{jnfu@bnu.edu.cn}

\author{Weikai Zong}
\affil{Department of Astronomy, Beijing Normal University, Beijing~100875, P.~R.~China}

\author{Jian-Ning Fu}
\affil{Department of Astronomy, Beijing Normal University, Beijing~100875, P.~R.~China}

\author{Peter De Cat}
\affil{Royal observatory of Belgium, Ringlaan 3, B-1180 Brussel, Belgium}

\author{Jianrong Shi}
\affil{Key Lab for Optical Astronomy, National Astronomical Observatories, Chinese Academy of Sciences, Beijing 100012, P.~R.~China}

\author{Ali Luo}
\affil{Key Lab for Optical Astronomy, National Astronomical Observatories, Chinese Academy of Sciences, Beijing 100012, P.~R.~China}

\author{Haotong Zhang}
\affil{Key Lab for Optical Astronomy, National Astronomical Observatories, Chinese Academy of Sciences, Beijing 100012, P.~R.~China}

\author{A. Frasca}
\affil{INAF---Osservatorio Astrfisico di Catania, Via S. Sofia 78, I-95123 Catania, Italy}

\author{C. J. Corbally}
\affil{Vatican Observatory Research Group, Steward Observatory, Tucson, AZ 85721-0065, USA}

\author{J. Molenda- \.Zakowicz}
\affil{Astronomical Institute of the University of Wroc\l{}aw, ul. Kopernika 11, 51-622 Wroc\l{}aw, Poland}

\author{G. Catanzaro}
\affil{INAF---Osservatorio Astrfisico di Catania, Via S. Sofia 78, I-95123 Catania, Italy}

\author{R. O. Gray}
\affil{Department of Physics and Astronomy, Appalachian State University, Boone, NC 28608, USA}

\author{Jiangtao Wang}
\affil{Department of Astronomy, Beijing Normal University, Beijing~100875, P.~R.~China}

\author{Yang Pan}
\affil{Department of Astronomy, Beijing Normal University, Beijing~100875, P.~R.~China}

\author{Anbing Ren}
\affil{Department of Astronomy, Beijing Normal University, Beijing~100875, P.~R.~China}

\author{Ruyuan Zhang}
\affil{Department of Astronomy, Beijing Normal University, Beijing~100875, P.~R.~China}

\author{Mengqi Jin}
\affil{Department of Astronomy, Beijing Normal University, Beijing~100875, P.~R.~China}

\author{Yue Wu}
\affil{Key Lab for Optical Astronomy, National Astronomical Observatories, Chinese Academy of Sciences, Beijing 100012, P.~R.~China}

\author{SuBo Dong}
\affil{Kavli Institute for Astronomy and Astrophysics, Peking University, Yi He Yuan Road 5, Hai Dian District, Beijing, 100871, P.~R.~China}

\author{Jiwei Xie}
\affil{School of Astronomy and Space Science, Nanjing University, Nanjing 210093, China}
\affil{Key Laboratory of Modern Astronomy and Astrophysics in Ministry of Education, Nanjing University, Nanjing 210093, China}

\author{Wei Zhang}
\affil{Key Lab for Optical Astronomy, National Astronomical Observatories, Chinese Academy of Sciences, Beijing 100012, P.~R.~China}

\author{Yonghui Hou}
\affil{Nanjing Institute of Astronomical Optics \& Technology, National Astronomical Observatories, Chinese Academy of Sciences, Nanjing~210042, P.~R.~China}

\author{LAMOST-{\sl Kepler} collaboration}


\begin{abstract}

The LAMOST-{\sl Kepler} (LK-) project was initiated to use the Large Sky Area Multi-Object Fiber Spectroscopic Telescope (LAMOST) to make spectroscopic follow-up observations for the targets in the field of the {\sl Kepler} mission. The {\sl Kepler} field is divided into 14 subfields that are adapted to the LAMOST circular field with {a} diameter of 5 degrees. During the regular survey phase of LAMOST, the LK-project took data from 2012 June to 2017 June {and} covered all the 14 subfields at least twice. In particular, we describe in this paper the second Data Release of the LK-project, including all spectra acquired through 2015 May to 2017 June together with the first round observations of {the} LK-project from 2012 June to 2014 September. The LK-project now counts 227\,870 spectra of 156\,390 stars, {among which we have derived atmospheric parameters ($\log g$, $T_\mathrm{eff}$ and [Fe/H]) and heliocentric radial velocity ($RV$) for 173\,971 spectra of 126\,172 stars. These parameters were obtained with the} most recent version of the LAMOST Stellar Parameter Pipeline v\,2.9.7. {Nearly one half}, namely { 76\,283} targets, are observed both by LAMOST and {\sl Kepler} telescopes. These spectra, establishing a {large} spectroscopy library, will be useful for the entire astronomical community, particularly for planetary science and stellar variability on {\sl Kepler} targets.

\end{abstract}


\section{Introduction}

The {\sl Kepler} satellite was launched on 2009 March 7th by {\sl NASA} with the aim of searching for Earth-sized planets around Solar-like stars \citep{2010Sci...327..977B}. Before its second of four reaction wheels on board failed on 2013 May 11th, {\sl Kepler} had been continuously monitoring about 200\,000 stars together within a 105 deg$^2$ field in the constellations Cygnus and Lyrae region for a period of $\sim 4$\,yr. These unprecedented high-quality photometric data are  goldmines for the field of asteroseismology \citep{gilliand10}, as well as for many other science cases \citep[see, e.g., eclipsing binaries in ][]{2011AJ....141...83P}. Nevertheless, to uncover {in-depth} physics for interesting targets with {\sl Kepler} photometry, it is crucial that the input atmospheric parameters of stars are available beforehand, such as {for} the successful asteroseismology depending somewhat on the effective temperature ($T_\mathrm{eff}$), surface gravity ($\log g$) and metallicity ([$M/H$]) measured first from spectroscopy, 
{that could reduce the size of parameter space to find the optimal seismic models} 
\citep[see, e.g., ][]{2007A&ARv..14..217C,2011A&A...530A...3C}. The {\sl Kepler Input Catalog} \citep[KIC;][]{2011AJ....142..112B} provides stellar parameters of potential interesting targets which are derived mainly from {photometry in the SDSS-like photometric bands \citep{2010AJ....139.1628D}} but the precision of KIC is not high enough for asteroseismic modeling, particular for hot and peculiar stars \citep{2011MNRAS.412.1210M,2012AJ....143..101M,2014ApJS..211....2H}. There are many additional endeavors to improve the precision of atmosphere parameters for seismic aims with ground-based spectroscopic data \citep[see, e.g.,][]{2010AN....331..993U,2012A&A...543A.160T,2015MNRAS.450.2764N}. It is inevitable that these projects are not allocated sufficient observation time to fully cover all $\sim200\,000$ {\sl Kepler} targets, except telescope installed with a large number of {fibers}. In addition, the spectroscopic data from different instruments may suffer systematic uncertainties. {Recently, spectroscopic follow-up works in the {\sl Kepler} field for asteroseismolgy have also been performed by the APOKASC with the multiple fibers of SDSS telescope, which is a part {of a} survey of APOGEE \citep{2017ApJS..233...23S,2017AJ....154...94M,2018arXiv180409983P}.}

The Large Sky Area Multi-Object Fiber Spectroscopic Telescope (LAMOST, also called Gou Shoujing Telescope) is an ideal instrument for follow-up spectroscopic observations on {\sl Kepler} stars. It combines a large aperture (3.6--4.9 m) with a wide field of view (circular FOV with the diameter of 5$^\circ$) and is equipped with 4000 {fibers} {at} its focus \citep{1996ApOpt..35.5155W,1998SPIE.3352..839X}. LAMOST spectra have a low resolution $R\sim1800$ and cover the wavelength range from 370\,nm to 900\,nm \citep[see details in][]{2012RAA....12.1197C,2012RAA....12..723Z}. {To take advantage of the ability of LAMOST to acquire many spectra at a time over a large field of view}, the LAMOST-{\sl Kepler} (LK) project was initiated with {the} scientific goal to observe as many objects in the {\sl Kepler} FOV as possible from the test and pilot survey phase of LAMOST onwards \citep[][hereafter Paper\,\RNum{1}]{2015ApJS..220...19D}. This strategy provides a homogeneous determination of both the atmosphere parameters (i.e., the surface gravity $\log g$, the effective temperature $T_\mathrm{eff}$ and the metallicity [Fe/H]) and the spectral classification of the observed objects. The low resolution spectra can also be used to estimate the radial velocity ($RV$) and the projected rotational velocity ($v\sin i$) for the rapid rotation stars.

During the first round of observation, from 2011 May to 2014 September, {the} LK-project had obtained 101\,086 low resolution spectra, covering about $21\%$ of 200\,000 {\sl Kepler} stars (Paper\,\RNum{1}). These spectra had been independently analyzed by three different groups: (i) the ``Asian team'' \citep{2016ApJS..225...28R} performed a statistical analysis of the stellar parameters based on the LAMOST stellar parameter pipeline \citep[LASP;][]{2011A&A...525A..71W,2014IAUS..306..340W,2015RAA....15.1095L}; (ii) the ``European team'' \citep{2016A&A...594A..39F} determined the stellar parameters,  the spectral classification, and the activity indicators with an updated version of the code ROTFIT \citep{2003A&A...405..149F,2006A&A...454..301F}; (iii) the ``American team'' \citep{2016AJ....151...13G} developed the code MKCLASS which automatically classifies stellar spectra on the MK system independent of the stellar parameter determination \citep[see details in][]{2014AJ....147...80G}. The LAMOST-{\sl Kepler} spectra, {by} establishing a spectroscopic database, have {made an impact on many scientific areas}, particular for the research within the {\sl Kepler} community. These spectra cover the Lithium line at 670.8\,nm and were used to discover the first confirmed Li-rich core-helium-burning giant \citep{2014ApJ...784L..16S}. The {spectra} also cover the Ca\,\RNum{2}\,H and K lines at 396.85 and 393.37\,nm, respectively, which have been used to measure the chromospheric activity on the stellar surface for 5648 solar-like stars within the {\sl Kepler} field based on LK spectra \citep{2016NatCo...711058K}. With the help of LAMOST spectra, the metallicity of the open cluster NGC\,6866 has been determined to be similar with the solar value \citep{2015MNRAS.453.1095B}. These spectra are also very useful for individual asteroseismic cases such as the red giant star KIC\,5689820 \citep{2014A&A...564A..27D} and the main-sequence A-type pulsating star KIC\,7917485 \citep{2016ApJ...827L..17M}. With the metallicity derived from the LAMOST-{\sl Kepler} spectra, \citet{2017MNRAS.465.2662S} found evidence that the incidence of pulsations in Am stars decreases with increasing {metallicity}. The LAMOST metallicity {has also play an important role} in the research of {\sl Kepler} exoplanets where {an excess of hot rocky {\it Kepler} planets was} found to be preferentially orbiting {around} metal-rich stars \citep{2016AJ....152..187M} 
{and a new population of short-period ($P<10$ d) exoplanets with sizes $R_p = 2-6 R_\earth$  that resemble hot Jupiters was discovered in preferentially occurring around metal-rich hosts and in single-transiting systems.} \citep{2018PNAS..115..266D}. With the atmospher{ic} parameters derived from LK-project  spectra, stars {hosting multiple transiting} planets are found with typically near circular or coplanar orbits which is similar to our Solar system \citep{2016PNAS..11311431X} {while {\sl Kepler} singles are on average eccentric ($\bar{e} \approx 0.3$). The LK sample is also helpful for investigating the intrinsic architecture of {\it Kepler} planetary systems, showing that the frequency of {\it Kepler}-like planetary systems is about $30\%$ \citep{2018ApJ...860..101Z}}. In addition, the LK-project is a part of {a} large spectroscopic survey for Galactic archaeology \citep[see, e.g.,][]{2017MNRAS.464.3657X}.

Given the importance of these spectra, we report in this paper the second round of spectroscopic observations in the {\sl Kepler} field with {the} LAMOST telescope from 2015 May to 2017 June. We also include the spectra obtained from the first round observations since they are now extracted and analyzed with the {most up-to-date} pipelines. The structure of this paper is organized as follows: we first describe the details of {the} observation{s} for this second round in Section~2, while the database of this version of spectra acquired is given in Section~3, with the discussion followed in Section~4, and {we} end with a brief summary. 

\section{Observations}
\begin{deluxetable}{cccrr}
\centering
\tablecaption{General contents of the LK-project observations during the regular survey phase from 2012 to 2017. \label{t1}}
\tablehead{
\colhead{Year} & \colhead{LK field}& \colhead{Plate}& \colhead{Spectra}& \colhead{Parameter}}
\startdata
2012 & 3  &   7  & 17659  & 11682    \\
2013 & 6  &  14  & 39309  & 28115    \\
2014 & 7  &  14  & 38516  & 29351    \\
2015 & 11 &  32  & 97247  & {81381}    \\
2017 & 6  &  16  & 35139  & 23442    \\
\hline
Total  &   &     & 227\,870  & 173\,971 \\
Unique &   &     & 104\,887  &  89\,570 \\
2$\times$ &  &   &  37\,482  &  28\,077 \\
3$\times$ &  &   &  10\,552  &   6\,613 \\
4$\times$ &  &   &   2\,293  &   1\,429 \\
+5$\times$ & &   &   1\,176  &    483   \\
\enddata
\tablecomments{The number of multiple revisited targets depends on the criteria one choose when performing cross-identification.}
\end{deluxetable}

As LAMOST has a focal plane of 5 degrees in diameter, a minimum of 14 circular LAMOST pointings (or footprints) is needed to cover the 105 square degrees of the {\sl Kepler} field. We refer to these individual footprints as LK-fields and the details of classification of each field can be seen in Paper\,\RNum{1}. 
The observed plates are called ``V-plates'', ``B-plates'', ``M-plates'', and ``F-plates'' for targets with the magnitudes of $9< r \leqslant 14$, $14< r \leqslant 16.3$, $16.3< r \leqslant 17.8$, and $17.8< r \leqslant 18.5$, respectively. In the first round of LK-project observations, we focused on stars which were included on the list of possible targets of the KASC\footnote{http://www.kasoc.dk} (Paper\,\RNum{1}).
{The targets for each plate were selected from a prioritized target list for the second round of observations, compiled from two main types of objects from high to low priority. (1) The 168,151 objects observed by the {\it Kepler} mission for which no high-quality LAMOST spectrum without potential issues (no entries in column\,14 of Table\,4 or column\,16 of Table\,5 of Paper\,I) was available after the first round of observations were considered as the new ``science targets''. Within this category, the seismology targets received a higher priority than the planet-search targets. (2) The new ``field stars'' consisted of three kinds of objects in order of decreasing priority: (a) the 31\,567 {\it Kepler} targets already having a clean high-quality LAMOST spectrum (to allow multi-epoch observations for quality assessment and variability studies), (b) the objects in the KIC that were not observed by {\it Kepler}, and (c) other stars in the fields-of-view with $V < 18$ mag from the USNO-B catalog \citep{2003AJ....125..984M}} Given the constraints of both the pointing ability of the LAMOST and the visibility of the {\sl Kepler} field,  observations can only be done from late-May to mid-October with a maximal observation window of four hours. In addition, LAMOST is closed during about two months per year in summer due to the Monsoon. Therefore, it takes three or four seasons for LAMOST to have a full coverage of the whole {\sl Kepler} field, considering the weather conditions as well. Table\,\ref{t1} lists the general information of observations made in each year through the regular survey phase of LAMOST from 2012 to 2017. The {multiple} visited targets are counted if spectra were observed with R.A. and DEC. within 3.7 arcsec\footnote{The value is adopted on the basis of that the fiber pointing precision is 0.4 arcsec and the diameter of the fiber is 3.3 arcsec.} and magnitude difference less than 0.01 mag. The LK-project {has been} carried out for two rounds of observations, meaning that all the 14 subfields have been observed at least once in each round. We note that the observations during the phase of the pilot survey of LAMOST {have not been processed with the newest version of the pipeline (c.f. Section~3.1) and }
are not considered here. {It only concerns a} small amount of data.

The details of observations of the first round of the LK-project from 2012 to 2014 are given in Paper\,\RNum{1}. Here we focus on the second round of observations of the LK-project, which began on 2015 May 29 and ended on 2017 June 15. A total of 48 plates were observed on 24  observation nights: 32 plates on 18 nights in 2015 and 16 plates on 6 nights 2017. Table\,\ref{T:LKOBS} lists the overview information of the observed plates for the second round of the LK-project. The very bright ``V-plate{s}'' were mainly observed as well as a small fraction of bright ``B-plate{s}'', with typical exposure times of $3\times10$ minutes and $3\times25$ minutes, respectively, depending on the observation conditions. 

\startlongtable
\begin{deluxetable*}{cccccccccc}
\centering
\tablecaption{Overview of the observed plates for the second round of the LK-project from 2015 to 2017. \label{T:LKOBS}}
\tablehead{
\colhead{PlateID} & \colhead{LK-field}& \colhead{PlanID}& \colhead{R.A. (2000)}& \colhead{DEC. (2000)}&\colhead{Date} & \colhead{Seeing}& \colhead{Exposure time} \\ \colhead{ }       & \colhead{ }       & \colhead{ }     & \colhead{ }          & \colhead{ }           & \colhead{ }   & \colhead{(arcsec) }     & \colhead{(s)} & 
}
\startdata
3506   &   LK14   &   KP192323N501616V04   &   19:23:23.787   &   +50:16:16.64   &   2015-05-29   &   2.40   &   600$\times$3     \\ 
3507   &   LK14   &   KP192323N501616V05   &   19:23:23.787   &   +50:16:16.64   &   2015-05-29   &   2.50   &   600$\times$2     \\ 
3509   &   LK11   &   KP190651N485531V02   &   19:06:51.499   &   +48:55:31.75   &   2015-05-30   &   3.00   &   600$\times$3     \\ 
3510   &   LK11   &   KP190651N485531V03   &   19:06:51.499   &   +48:55:31.75   &   2015-05-30   &   3.20   &   600$\times$3     \\ 
3511   &   LK11   &   KP190651N485531V04   &   19:06:51.499   &   +48:55:31.75   &   2015-05-30   &   3.50   &   600     \\ 
3538   &   LK14   &   KP192323N501616B01   &   19:23:23.787   &   +50:16:16.64   &   2015-09-13   &   4.00   &   1500$\times$3     \\ 
3542   &   LK10   &   KP192314N471144B02   &   19:23:14.829   &   +47:11:44.80   &   2015-09-14   &   3.20   &   1500$\times$3     \\ 
3544   &   LK11   &   KP190651N485531B02   &   19:06:51.499   &   +48:55:31.75   &   2015-09-15   &   2.40   &   1500$\times$3     \\ 
3546   &   LK06   &   KP194045N483045B01   &   19:40:45.382   &   +48:30:45.10   &   2015-09-16   &   3.20   &   1500$\times$3     \\ 
3551   &   LK08   &   KP195920N454621B01   &   19:59:20.424   &   +45:46:21.15   &   2015-09-18   &   3.00   &   1500$\times$3     \\ 
3552   &   LK08   &   KP195920N454621B02   &   19:59:20.424   &   +45:46:21.15   &   2015-09-18   &   3.50   &   1500+1356$^{a}$     \\ 
3599   &   LK02   &   KP193637N444141V03   &   19:36:37.977   &   +44:41:41.77   &   2015-09-21   &   3.00   &   600$\times$3     \\ 
3600   &   LK02   &   KP193637N444141V04   &   19:36:37.977   &   +44:41:41.77   &   2015-09-21   &   3.30   &   600$\times$3     \\ 
3601   &   LK02   &   KP193637N444141V05   &   19:36:37.977   &   +44:41:41.77   &   2015-09-21   &   3.90   &   600$\times$2     \\ 
3620   &   LK09   &   KP190808N440210V02   &   19:08:08.340   &   +44:02:10.88   &   2015-09-25   &   3.00   &   600$\times$3     \\ 
3621   &   LK09   &   KP190808N440210V03   &   19:08:08.340   &   +44:02:10.88   &   2015-09-25   &   3.40   &   600$\times$3     \\ 
3627   &   LK06   &   KP194045N483045V03   &   19:40:45.382   &   +48:30:45.10   &   2015-10-01   &   4.20   &   600$\times$3     \\ 
3628   &   LK06   &   KP194045N483045V04   &   19:40:45.382   &   +48:30:45.10   &   2015-10-01   &   4.30   &   600$\times$3     \\ 
3634   &   LK10   &   KP192314N471144V02   &   19:23:14.829   &   +47:11:44.80   &   2015-10-02   &   3.40   &   600$\times$3     \\ 
3635   &   LK10   &   KP192314N471144V03   &   19:23:14.829   &   +47:11:44.80   &   2015-10-02   &   3.80   &   600$\times$2+561$^{a}$     \\ 
3642   &   LK08   &   KP195920N454621V04   &   19:59:20.424   &   +45:46:21.15   &   2015-10-03   &   2.70   &   600$\times$3     \\ 
3643   &   LK08   &   KP195920N454621V05   &   19:59:20.424   &   +45:46:21.15   &   2015-10-03   &   3.10   &   600$\times$3     \\ 
3650   &   LK12   &   KP185031N425443V02   &   18:50:31.041   &   +42:54:43.72   &   2015-10-04   &   2.80   &   600$\times$3     \\ 
3658   &   LK13   &   KP185111N464417V04   &   18:51:11.993   &   +46:44:17.52   &   2015-10-06   &   4.00   &   600$\times$3     \\ 
3677   &   LK05   &   KP194918N413456V04   &   19:49:18.139   &   +41:34:56.86   &   2015-10-08   &   3.70   &   600$\times$3     \\ 
3678   &   LK05   &   KP194918N413456V05   &   19:49:18.139   &   +41:34:56.86   &   2015-10-08   &   4.70   &   600$\times$3     \\ 
3679   &   LK05   &   KP194918N413456V06   &   19:49:18.139   &   +41:34:56.86   &   2015-10-08   &   2.90   &   600+579$^{a}$     \\ 
3690   &   LK07   &   KP192102N424113V03   &   19:21:02.816   &   +42:41:13.07   &   2015-10-11   &   2.80   &   600$\times$3     \\ 
3691   &   LK07   &   KP192102N424113V04   &   19:21:02.816   &   +42:41:13.07   &   2015-10-11   &   3.20   &   600$\times$3     \\ 
3697   &   LK03   &   KP192409N391242V01   &   19:24:09.919   &   +39:12:42.00   &   2015-10-12   &   3.00   &   600$\times$3     \\ 
3698   &   LK03   &   KP192409N391242V02   &   19:24:09.919   &   +39:12:42.00   &   2015-10-12   &   3.60   &   600$\times$3     \\ 
3731   &   LK04   &   KP193708N401249V01   &   19:37:08.863   &   +40:12:49.63   &   2015-10-18   &   4.20   &   600$\times$3     \\ 
5764   &   LK01   &   KP190339N395439B01   &   19:03:39.258   &   +39:54:39.24   &   2017-06-03   &   2.50   &   1500$\times$3     \\ 
5765   &   LK01   &   KP190339N395439V03   &   19:03:39.258   &   +39:54:39.24   &   2017-06-03   &   3.60   &   600$\times$2     \\ 
5776   &   LK12   &   KP185031N425443V03   &   18:50:31.041   &   +42:54:43.72   &   2017-06-07   &   2.50   &   600$\times$3     \\ 
5777   &   LK12   &   KP185031N425443V04   &   18:50:31.041   &   +42:54:43.72   &   2017-06-07   &   2.60   &   600$\times$3     \\ 
5778   &   LK12   &   KP185031N425443V05   &   18:50:31.041   &   +42:54:43.72   &   2017-06-07   &   2.70   &   600$\times$4     \\ 
5798   &   LK04   &   KP193708N401249V02   &   19:37:08.863   &   +40:12:49.63   &   2017-06-12   &   3.20   &   600$\times$3     \\ 
5799   &   LK04   &   KP193708N401249V03   &   19:37:08.863   &   +40:12:49.63   &   2017-06-12   &   3.40   &   600$\times$3     \\ 
5805   &   LK03   &   KP192409N391242V03   &   19:24:09.919   &   +39:12:42.00   &   2017-06-13   &   3.50   &   600$\times$3     \\ 
5806   &   LK03   &   KP192409N391242V04   &   19:24:09.919   &   +39:12:42.00   &   2017-06-13   &   2.70   &   600$\times$3     \\ 
5807   &   LK03   &   KP192409N391242V05   &   19:24:09.919   &   +39:12:42.00   &   2017-06-13   &   2.90   &   600$\times$3     \\ 
5811   &   LK10   &   KP192314N471144V04   &   19:23:14.829   &   +47:11:44.80   &   2017-06-14   &   2.80   &   600$\times$3     \\ 
5812   &   LK10   &   KP192314N471144V05   &   19:23:14.829   &   +47:11:44.80   &   2017-06-14   &   2.90   &   600$\times$3     \\ 
5813   &   LK10   &   KP192314N471144V06   &   19:23:14.829   &   +47:11:44.80   &   2017-06-14   &   3.50   &   600$\times$3     \\ 
5816   &   LK09   &   KP190808N440210V04   &   19:08:08.340   &   +44:02:10.88   &   2017-06-15   &   3.30   &   600$\times$3     \\ 
5817   &   LK09   &   KP190808N440210V05   &   19:08:08.340   &   +44:02:10.88   &   2017-06-15   &   3.70   &   600$\times$3     \\ 
5818   &   LK09   &   KP190808N440210V06   &   19:08:08.340   &   +44:02:10.88   &   2017-06-15   &   4.00   &   600$\times$3    \\ 
\enddata
\tablecomments{For each plate, we give the sequence number of the plate (PlateID), the reference of the LK-field (LK\#), the plan identification number (PlanID), with the associated right ascension (R.A. (2000)) and declination (DEC. (2000)) of the central star on that plate, the observed date (Date), the air turbulence in Xinglong (seeing), and the time of each exposure (Exposure time). {\bf Note that the plate KP193708N401249V04 was observed on 2017-06-12 but it suffered from a bad weather condition and all spectra have a very low signal-to-noise. We therefore did not include that plate in this table.}}
\tablenotetext{a}{The {\bf unusual} exposure time happened {\bf because the plate moved out of the observable view of LAMOST}, two hours before and after its meridian, respectively.}

\end{deluxetable*}

\section{Spectra release}
\subsection{Data reduction process}
One of the main products of LAMOST provided to astronomers is the wavelength and flux-calibrated spectra, which are processed by an automatic data reduction and data analysis pipeline \citep[see details in][]{2012RAA....12.1243L,2015RAA....15.1095L}. The version v2.9.7 pipeline is used for the spectra obtained in the five-year regular survey of LAMOST from 2012 to 2017. After the quality evaluation of the observations {(e.g., seeing and cloud coverage) and telescope performance}, the raw CCD frames are fed to the 2D pipeline which uses procedures similar to those of SDSS \citep{2002AJ....123..485S}, {to produce calibrated 1D spectra}. The main tasks of the 2D pipeline include dark and bias subtraction, flat field correction, spectral extraction, sky subtraction, wavelength calibration, stacking sub-exposures and combining of different wavelength bands \citep[the detailed description can be found in Paper\,\RNum{1} and][]{2015RAA....15.1095L}. The 1D pipeline 
is aimed at the analysis of the LAMOST spectra and provides the classification of spectral type and the measure of RVs (for stars) or redshifts (for galaxies and quasi-stellar objects). This goal is reached with the help of a template-matching and line-recognition algorithm. The library of stellar template{s} was constructed based on the classification of about one million LAMOST DR1 stellar spectra \citep{2014AJ....147..101W}.

\subsection{Data release}
\begin{figure*}
\centering
\includegraphics[width=12cm]{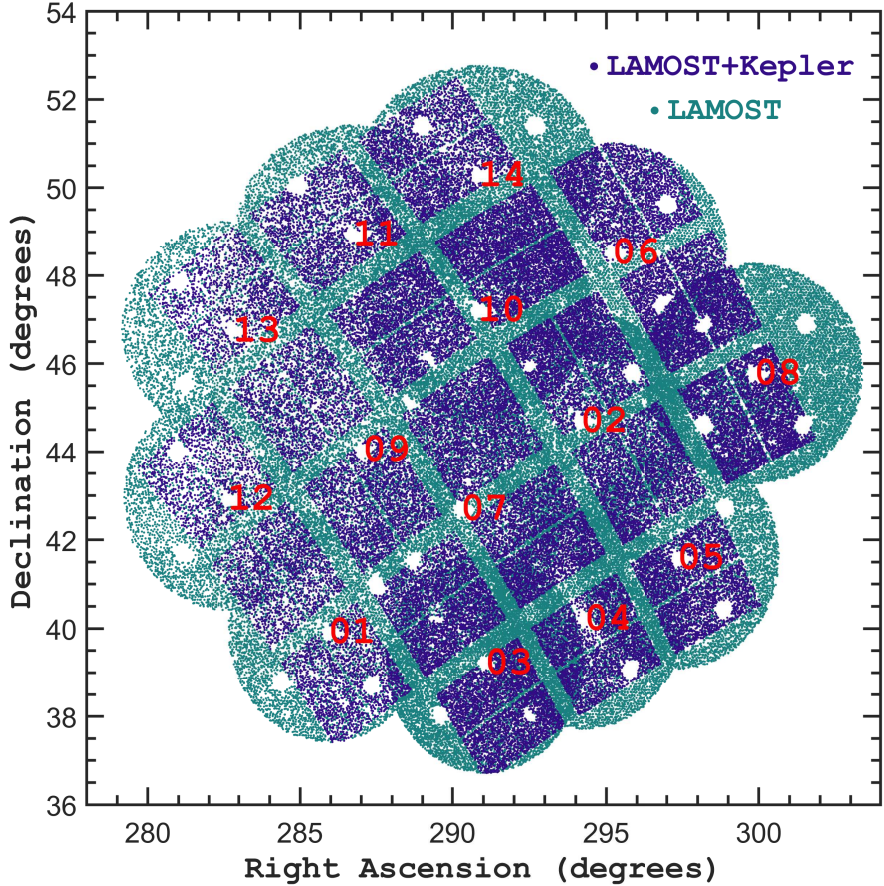}
\caption{Sky coverage of all targets observed by the LK project. The stars observed by LAMOST and with the {\sl Kepler} photometry are marked in Persian blue, others are in dark cyan. The numbers {mark} the central position of the 14 LK fields.
\label{f:rd}}
\end{figure*}
\begin{figure}
\centering
\includegraphics[width=8.5cm]{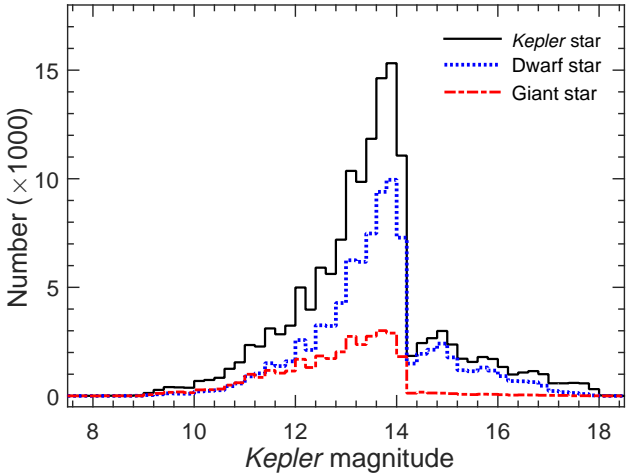}
\caption{The {\sl Kepler} magnitude ({\sl Kp}) distribution of the stars observed by LAMOST during the second round observations. Black solid, blue dotted and red dashed histograms represent the entire LK stars (not including the LAMOST standard stars and the stars without the available {\sl Kp} magnitude), the dwarfs, and the giants, respectively.  
\label{f:kp}}
\end{figure}

\begin{figure*}
\centering
\includegraphics[width=14cm]{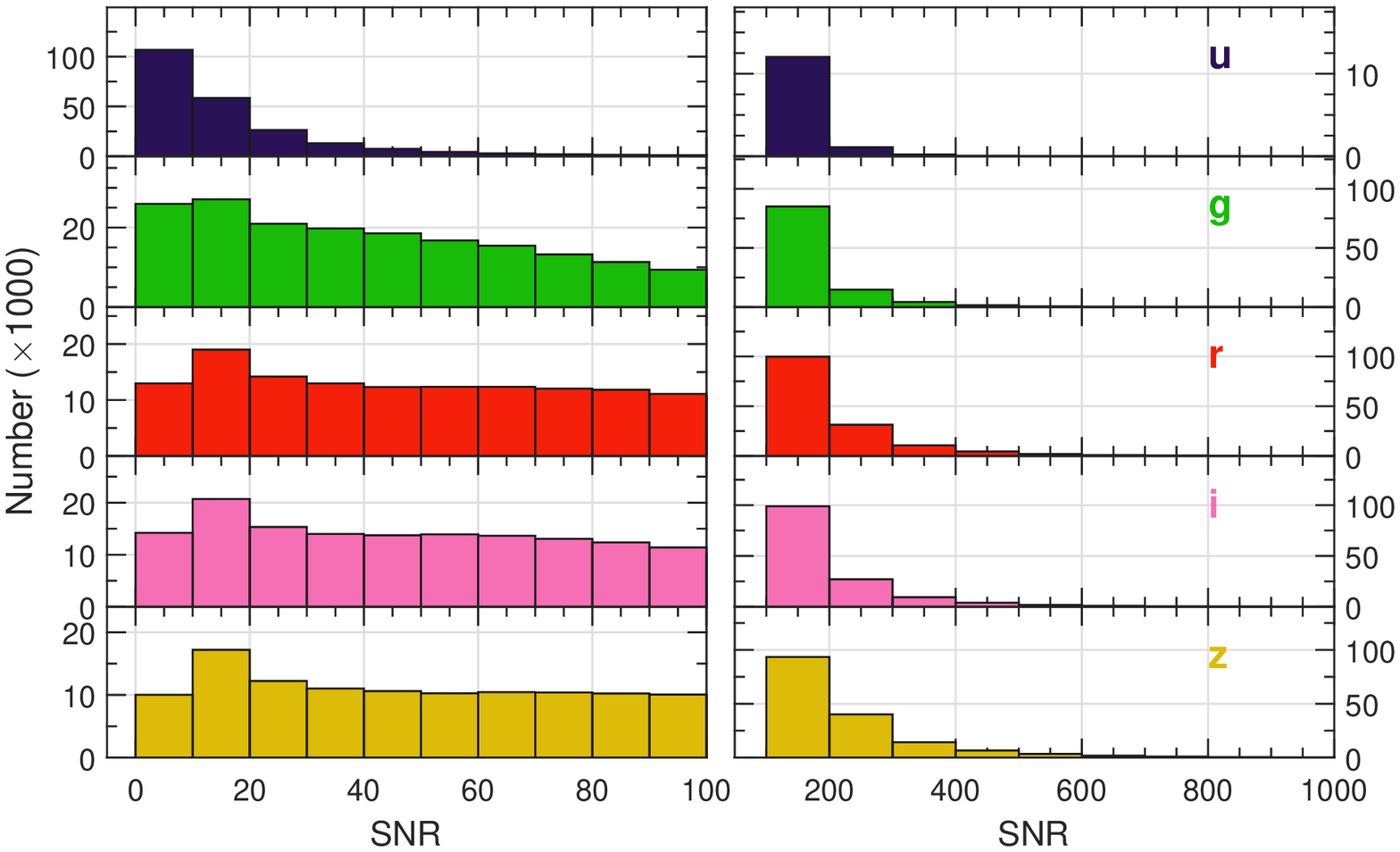}
\caption{The distributions of the S/N in the SDSS {\sl u}, {\sl g}, {\sl r}, {\sl i} and {\sl z} bands ( from the top down to the bottom panels) for the LK spectra. The left and right panels show the S/N range [0, 100] and [100, 1000], with bin size 10 and 100, respectively.
\label{figsnr}}
\end{figure*}

The calibrated spectra are released to the astronomical community at regular intervals, {typically once every year}, and {the pipelines are updated every one or two years,} during the regular surveys of LAMOST. After two rounds of observations of the LK-project, we have collected 227\,870 flux- and wavelength-calibrated, sky-subtracted low resolution ($R=1800$) spectra, among which 173\,971 ones have been used to calculate the stellar atmospheric parameters with the LASP pipelines. These data will be released to the public along with the fifth data release (DR5\footnote{http://dr5.lamost.org}) of all the LAMOST survey spectra in June 2019. They then can be freely downloaded from the official LAMOST website\footnote{www.lamost.org}. 

Figure\,\ref{f:rd} illustrates the spatial distribution of the stars observed within the LK-project from 2012 to 2017. The {spatial distribution of the} entire data set cover almost the entire area of the {\sl Kepler} field. There are 
{ 76\,283}\footnote{The number is obtained through cross-identification between the two datasets on the basis of coordinate separation of 3.0 arcsec. We note that the number would decrease when we use more strict constraints to perform cross-identification. For instrance, the number would be around 60\,000 if we would use a maximum coordinate separation of 0.4 arcsec and a magnitude difference of less than 0.01 mag.}
stars {which are both observed by LAMOST and have {\sl Kepler} photometry}, corresponding to a fraction of more than a third {of the full database of stars observed for the LK-project. We now have LAMOST data for { about 38.2\% of the $\sim200,000$ objects that have been observed during the {\sl Kepler} mission}. The central hole of each plate is reserved to a bright ($V<8$) star which is used for the active optics wavefront sensor {to shape the mirror getting rid of deformation {of the wavefront} from external influences. The coordinates of that star also defines the coordinates of the observed plate. Four guide stars ($V<17$) are observed by the guiding CCD cameras placed at the four corners of the LAMOST field of view.} Figure\,\ref{f:kp} shows the histogram of the {\sl Kepler} magnitude ({\sl K}p) distribution for the 146\,001 stars, including  dwarfs and giants. There are additional 10\,389 stars {that are not included among them because they} are the LAMOST standard stars or stars without {\sl K}p magnitude available from the DR5 catalog. Here we separate dwarfs from giants by their measured surface gravity with a boundary of $\log g = 3.5$. The histogram of Figure\,\ref{f:kp} clearly reveals that most targets fall into the range of {\sl K}p 11--14, with a small fraction of brighter and fainter targets. This histogram also depicts the observation strategy of the LK-project under different observation conditions. The observations concentrate on very bright plates (see Table\,\ref{T:LKOBS}, and Table~3 of Paper\,\RNum{1}), which can make full use of bright nights or nights with unfavorable weather conditions such as poor seeing or low atmospheric transparency. Consequently, a sharp cutoff appears at $K\rm{p}=14$ which is the boundary between the  $V$-plates and the $B$-plates. 

Figure\,\ref{figsnr} {displays the data quality showing the distributions of the} signal-to-noise ratios (S/Ns) in SDSS {\sl u\rm{,} \sl g\rm{,} \sl r\rm{,} \sl i\rm{, and} \sl z} bands. We note that the spectra with ${S/N_g}\geqslant6$ can be used to derive valid stellar parameters for type A, F, G and K stars with {the} LASP pipeline if they are obtained in dark nights (eight nights before and after the {new} moon). However, the criterion of ${S/N_g}$ {increases} to 15 for the spectra obtained in bright nights (all other nights except the three nights around the full moon). One can see more details of setting those criteria for calculation of atmospheric parameters in \citet{2015RAA....15.1095L}. After two  rounds of LK-project observations, {we observed merely 29\,505, 40\,916, and 53\,047 spectra with ${S/N_g}\leqslant6$, ${S/N_g}\leqslant10$ and ${S/N_g}\leqslant15$, {observed during both bright and} dark nights, which correspond to percentages of $\sim 12.95\%$, $\sim 17.96\%$ and $\sim 23.28\%$, respectively}.

\begin{sidewaystable}[h!] 
\startlongtable
\begin{deluxetable*}{ccccccccccccccccc}
\fontsize{1}{2}
\hspace{-2cm}
\centering
\tablecaption{Database of the LAMOST spectra obtained for the LK-project from 2012 to 2017. \label{T:LKSPT}}
\tablehead{
\colhead{(1)}  & \colhead{(2)} & \colhead{(3)} & \colhead{(4)} & \colhead{(5)} & \colhead{(6)} &\colhead{(7)}& \colhead{(8)} \\
\colhead{Obsid$^a$}  & \colhead{KIC$^b$} & \colhead{Tcomment} & \colhead{R.A. (2000)} & \colhead{DEC. (2000)} & \colhead{S/N$_g$} &\colhead{Mag}& \colhead{Subclass} \\
\colhead{(9)}  & \colhead{(10)} & \colhead{(11)} & \colhead{(12)} & \colhead{(13)} & \colhead{(14)} &\colhead{(15)}& \colhead{(16)} \\
\colhead{$T_\mathrm{eff}$} & \colhead{$\log g$} & \colhead{[Fe/H]} & \colhead{$RV$} & \colhead{yyyy-mm-ddThh:mm:ss.ss} & \colhead{$\Delta$d} & \colhead{KO} & \colhead{Filename} \\
\colhead{(K)} 
& \colhead{(dex)} & \colhead{(dex)} & \colhead{(km/s)} & \colhead{} & \colhead{(arcsec)} 
}
\tiny
\startdata
...       & ...        & ...             &   ...          &   ...         &   ...  &   ...  &   ...  \\
..        &   ...      &   ...           &   ...          &   ...         & ...    &   ...  &  ...   \\
580608079  &  3958615  &  kplr003958615  &  292.6349200   &  39.02463200  & 111.19 &  12.80 &  G5  \\
4272.48$\pm$24.10 &  2.159$\pm$0.040     & -0.399$\pm$0.022  & -30.37$\pm$3.46 & 2017-06-14T02:06:16.0 &  0.028878 &  Y &  spec-57918-KP192409N391242V04\_sp08-079.fits.gz \\
580608081  & 3958237   & kplr003958237   &  292.5281700   &   39.08282100 & 83.08  & 13.71  & G5  \\
4773.65$\pm$29.01 &  2.509$\pm$0.048     &  -0.170$\pm$0.027 &  -8.53$\pm$3.65 &  2017-06-14T02:06:16.0  &  0.029562 &  Y &  spec-57918-KP192409N391242V04\_sp08-081.fits.gz \\
580608082  & 3957804  & 1275-11519892    &   292.4251750  &  39.04117800  & 30.22  & 13.20  & G3  \\
5828.24$\pm$88.58 &  4.174$\pm$0.145 &  0.201$\pm$0.086 &  -43.62$\pm$5.91 &  2017-06-14T02:06:16.0 &  2.88  & Y  & spec-57918-KP192409N391242V04\_sp08-082.fits.gz \\
580608084  &  3958877  &  kplr003958877  &      292.7043500   &    39.02043200 &  125.91  &   13.23 &   G2  \\
5816.16$\pm$18.31  &  4.028$\pm$0.030  &  0.119$\pm$0.017  &  -21.41$\pm$4.27 &   2017-06-14T02:06:16.0 &   0.027528 &   Y &   spec-57918-KP192409N391242V04\_sp08-084.fits.gz \\
580608086  & 3958406  &  kplr003958406    &   292.5780000   &   39.03916900 & 38.67  &  13.92 &  G3  \\
5883.81$\pm$61.31 &  4.449$\pm$0.101 &  0.103$\pm$0.059 &  -29.80$\pm$6.01 &  2017-06-14T02:06:16.0 &  0.02958  & Y  & spec-57918-KP192409N391242V04\_sp08-086.fits.gz\\
...       & ...        & ...             &   ...          &   ...         &   ...  &   ...  &   ...  \\
..        &   ...      &   ...           &   ...          &   ...         & ...    &   ...  &  ...   \\
\enddata

\label{t3}
\end{deluxetable*}
{
\tablecomments{The entire table can be downloaded through \sl dr5.lamost.org/doc/vac.}
\tablenotetext{a}{Obsid is the "fingerprint" of that spectrum. One can keep the obsid if they want to get more information of that spectrum. It can be easily accessed though the official LAMOST website searching engine: dr5.lamost.org/search} 
\tablenotetext{b}{One should be careful when using the KIC identification if it has a large separation $\Delta d$ or conficts with its tcomment (input target). }
}
\end{sidewaystable}

Table\,\ref{T:LKSPT} lists the full catalog of the constructed database that is available for the entire LK-project. This table contains the following columns: 
\begin{itemize}
\item[] Column\,1: Obsid is the unique identification number ID of the observed 1D spectrum.
\item[] Column\,2: the cross-identification of the target in {\sl Kepler} input catalog within the coordinate separation of three arcsec\footnote{{ We note that the criteria are different for cross-identification and self-identification (c.f. Section~2)}. The nearest star will be chosen if more than one stars were identified.}.
{\item[] Column\,3: the input target ID from KIC, SDSS, UCAC4, PANSTAR or another catalogue.}
\item[] Column\,{4}: the input right ascension (epoch 2000.0) of the fiber pointed to, in unit of degree.
\item[] Column\,{5}: the input declination (epoch 2000.0) of the fiber pointed to, in unit of degree.
\item[] Column\,{6}: the S/N$_g$ is the signal-to-noise ratio of the spectra in SDSS {\sl g} band which gives the quality estimation of the spectrum.
\item[] Column\,{7}: the {\sl Kepler} magnitude ({\sl K}p, if available, {otherwise the magnitude in a specific filter will be given}).
\item[] Column\,{8}: Subclass is spectral type of the target calculated by the 1D pipeline if they are stars of A, F, G, K and M type.
\item[] Column\,{9}: the effective temperature with the associated error given by LASP pipeline ($T_\mathrm{eff}$).
\item[] Column\,{10}: the surface gravity with the associated error given by LASP pipeline ($\log g$).
\item[] Column\,1{1}: the metallicity with the associated error given by LASP pipeline.
\item[] Column\,1{2}: the radial velocity with the associated error given by LASP pipeline.
\item[] Column\,1{3}: the {(local)} observation { time} of the spectrum.
\item[] Column\,1{4}: the coordinate difference between the observed coordinates of LAMOST and the cross-identification from KIC coordinates (in arcsec).
\item[] Column\,1{5}: the {availability of} {\sl Kepler} {photometry} ("Y" for yes and "N" for no).
{\item[] Column\,16: the file name of the LAMOST 1D fits file.
}
\end{itemize}

\subsection{Charaterization of atmospheric parameters}

\begin{figure*}
\centering
\includegraphics[width=12cm]{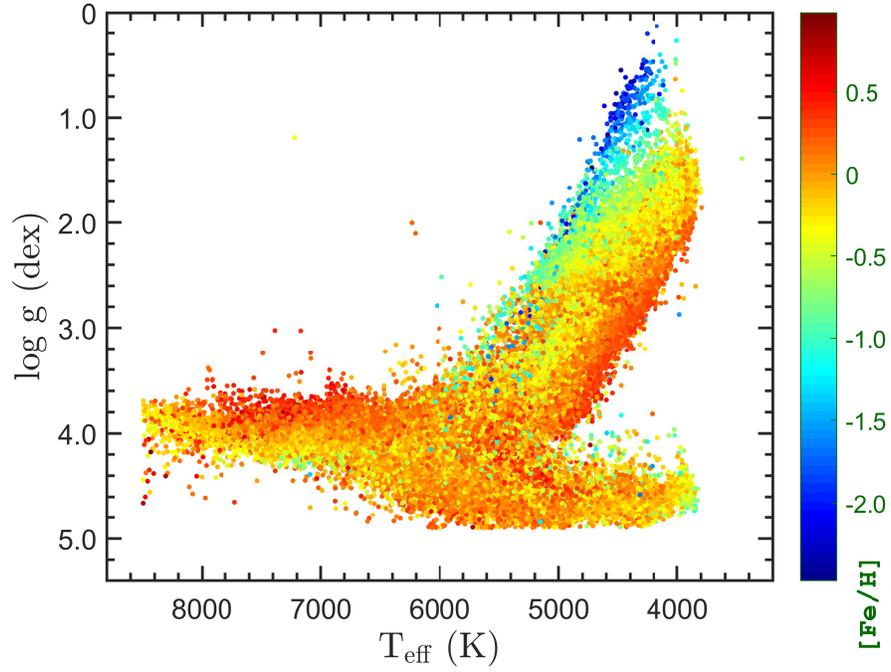}
\caption{The Kiel diagram ($\log g$ v.s. $T_\mathrm{eff}$) of the qualified LK spectra. The parameters are derived from LASP pipeline. Note that different colors indicate different values of metallicity [Fe/H].}
\label{swan}
\end{figure*}
Among these 227\,870 spectra, we provide stellar parameters for 173\,971 spectra with the LASP pipeline, as the spectra quality is high enough to achieve an accurate and reliable determination of parameters for the AFGK-types of stars \citep[with S/N$_g \geqslant 6$ and S/N$_g\geqslant15$ of spectra observed in dark and bright nights, respectively, see details in ][]{2015RAA....15.1095L}. We note that 51\,503 targets have been observed more than once (16$\times$ for the most observed stars) and the total number of stars {is} 156\,390. { There are 65\,529 {\sl Kepler} stars with LASP atmospheric parameters among the 76\,283 stars in common. We now have LASP parameters for about 32.8\% of the objects that have been observed during the {\sl Kepler} mission.} Figure\,\ref{swan} shows the {$\log g$ v.s. $T_\mathrm{eff}$ plane (also called Kiel diagram) for the spectra with sufficient quality.} The distribution of targets differs slightly but not significantly, which therefore is not provided here again. 
The surface gravities are found mainly between 5 and 1~dex, while the effective temperature is mainly in the range of [4000, 8000]~K. Most stars have metallicities near to the value of the Sun. We clearly see that most of them locate in either the main sequence or the red giant branch. {We note that the giant branch correctly displaces towards higher temperatures as the metallicity decreases.}

\begin{figure*}
\centering
\includegraphics[width=8.5cm]{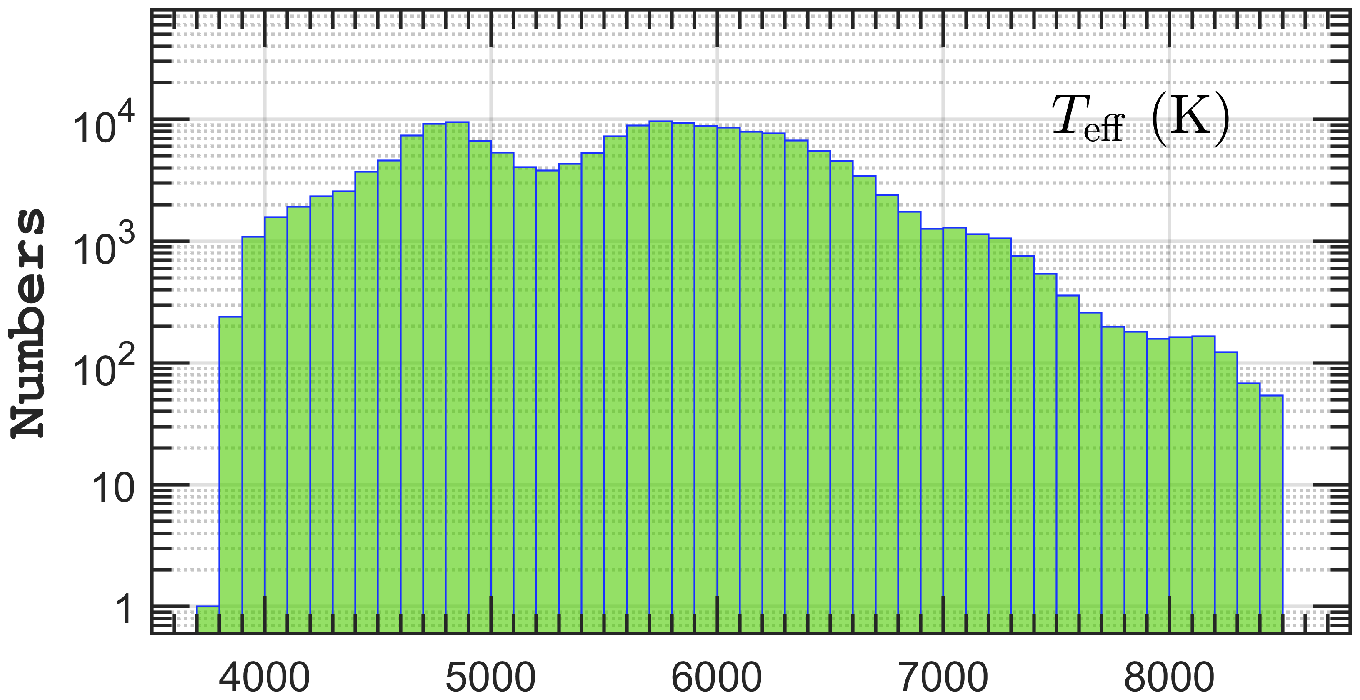}
\includegraphics[width=8.5cm]{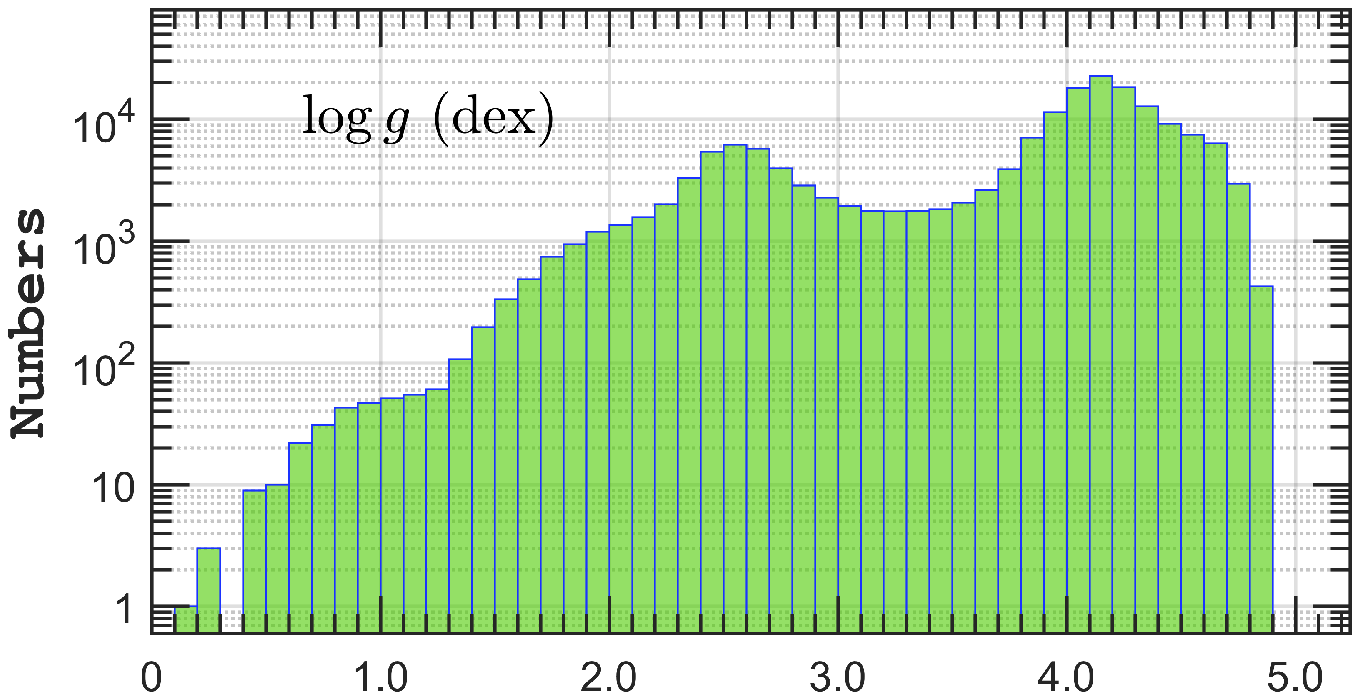}\\
\includegraphics[width=8.5cm]{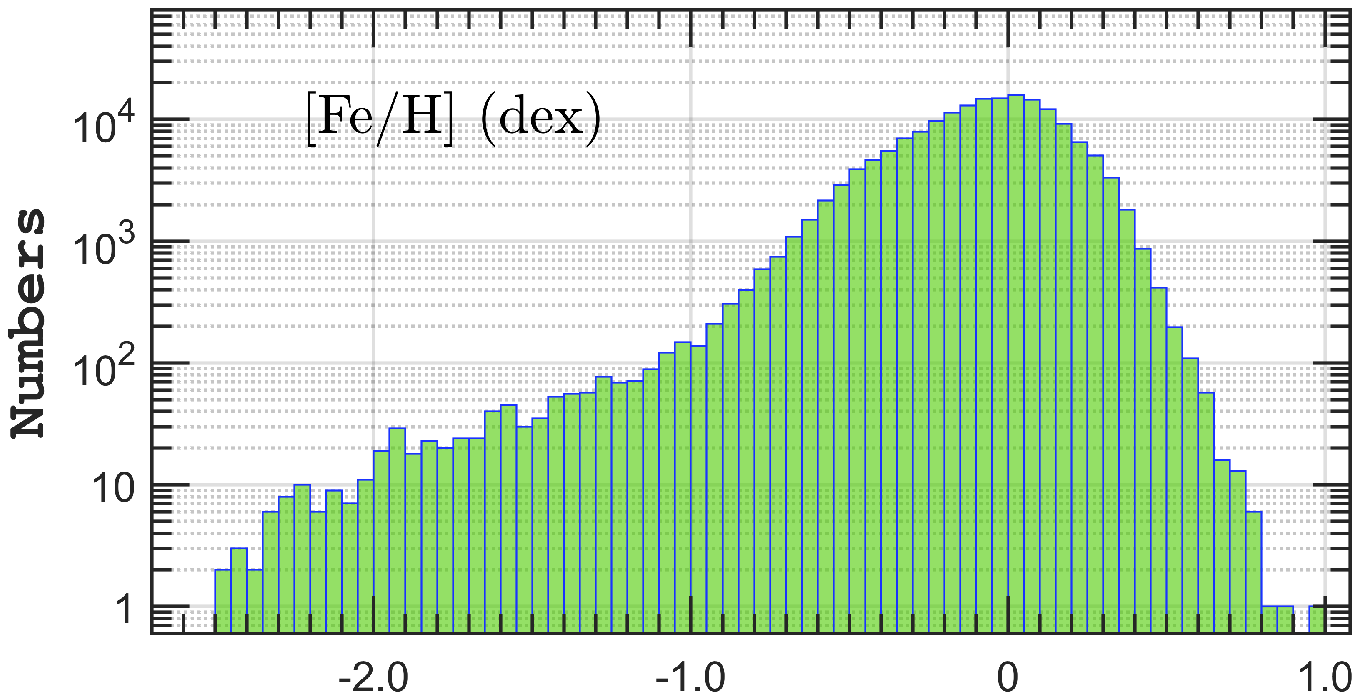}
\includegraphics[width=8.5cm]{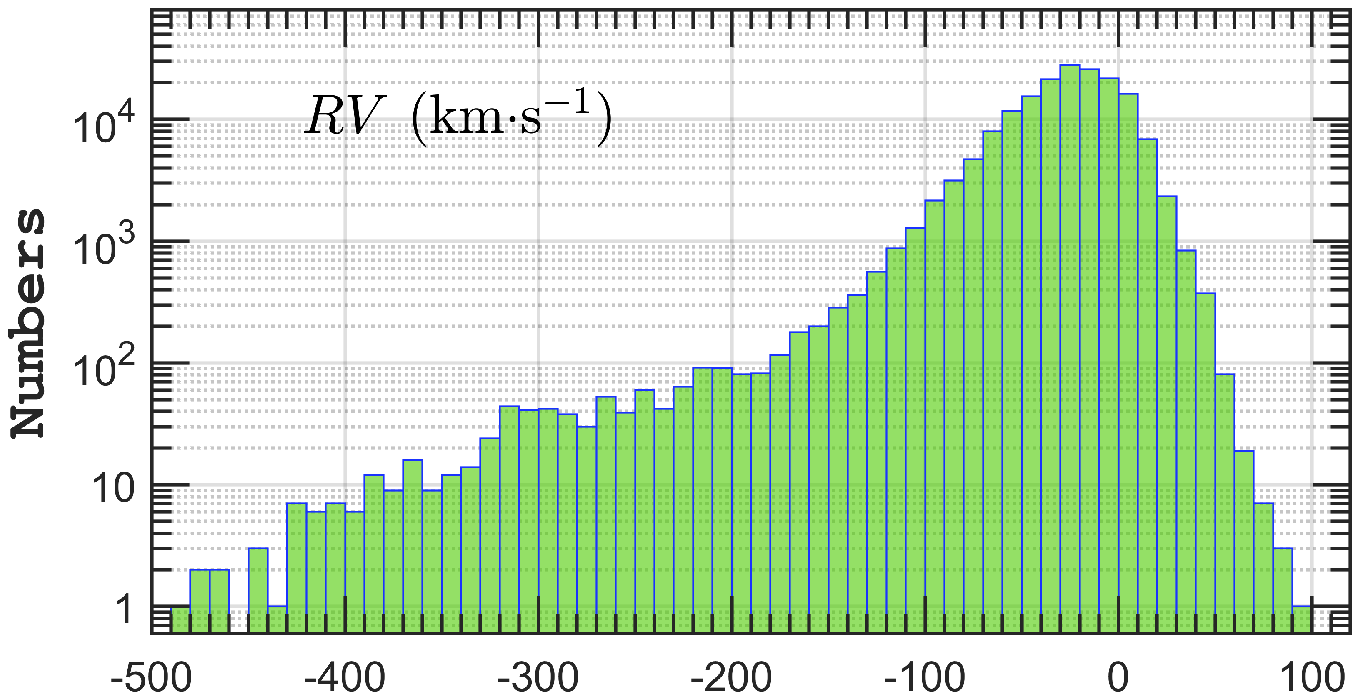}
\caption{Histogram distribution of atmospheric parameters derived from 173\,971 spectra. Top-left panel: the effective temperature $T_\mathrm{eff}$ (K, with bin of 100\,K); top-right: the surface gravity $\log g$ (dex, with bin of 0.1); bottom-left: the metallicity [Fe/H] (dex, with bin of 0.05); and bottom-right: the radial velocity $RV$ (km/s, with bin of 10\,km/s). }
\label{fours}
\end{figure*}

Figure\,\ref{fours} shows the histograms of the stellar parameters $T_\mathrm{eff}$,  $\log g$, [Fe/H] and the radial velocity. The distribution of effective temperature, $T_\mathrm{eff}$, {shows} a structure where two distinct peaks {are} found around 4700 and 5700~K, which are the projection of the giants and the main sequence stars from Figure\,\ref{swan}, respectively. The surface gravity, $\log g$, also presents two distinct peaks near 2.5 and 4.2 which, again, correspond to the projection of the giant and the main sequence stars, respectively. The metallicity {spans} the range from $\sim-2.0$ to 0.7~dex and peaks around the solar metallicity, which suggests that most stars have a nearly solar  metallicity. There are 1\,114 {stars with [Fe/H] $<-1.0$ and 65 stars with [Fe/H] $<-2.0$}, which we consider as candidate metal-poor stars and very metal-poor stars, respectively. They could be the fossils of the early generation of stars and can provide fundamental information on the chemical abundances, formation and evolution of the early evolutionary stage of the Galaxy \citep[see, e.g.,][]{2015ApJ...798..110L,1995AJ....109.2757M}. Most stars are found with radial velocities around $-20$\,km/s. However, in the long tail of the $RV$ distribution, we find 29 and 216 stars with $RV<-400$\,km/s and $-300$\,km/s, which are candidates of high-velocity stars \citep[see, e.g.,][]{2014ApJ...789L...2Z}. {The properties of these} stars can be used to determine their original nature and formation mechanisms and {aid the study} of the structural properties of the Galaxy \citep[see, e.g.,][]{2009MNRAS.396..570G}. 

\section{Discussion}
After completing the entire two-round observations of the LK-project, we have collected {227\,870} low resolution spectra of {156\,390} stars and provided atmospheric parameters with {173\,971} spectra of {126\,172} stars. {We remark that the low resolution spectra in the first round observations were mainly produced from the LASP pipeline with version~2.7.5. The parameter comparison from those two different versions is reported {i}n the data release note of DR5 website\footnote{http://dr5.lamost.org/doc/release-note}, where it is shown that the atmospheric parameters $T_\mathrm{eff}$, $\log g$, [Fe/H] and $RV$ are within their errors for the stars in common between DR3 and DR5. We note that, for stars within the LK-project, the parameters are also essentially the same, as displayed by  the similarity of Figure~6 in Paper\,\RNum{1} with Figures\,\ref{swan} and \ref{fours} here. The slight difference between the two Kiel diagrams is {mainly the result of {excluding} from the analysis the parameters of hot stars (OB-types) because the LASP pipeline is optimized for AFGK-type stars and could lead to untrustable results in other cases.} Nevertheless, as the size of {the} sample has {increased}, we observe more interesting targets such as candidates of very low metallicity stars and high-velocity stars (see histograms of Figure~6 of Paper\,\RNum{1} and our Figure\,\ref{fours}).} 

For the spectral database {of only} the first round of the LK-project, \citet{2016ApJS..225...28R} calibrated the atmospheric parameters derived from LASP by comparing them with the values measured from either high-resolution spectroscopy or asteroseismology for the stars in common with the catalog of \citet{2014ApJS..211....2H}. {A similar comparison was also performed by \citet{2014ApJ...789L...3D,2018PNAS..115..266D} and \citet{2016AJ....152....6W}.} \citet{2016ApJS..225...28R} provide the external errors of $T_\mathrm{eff}$, $\log g$ and [Fe/H] for giants and dwarfs, respectively. 
The internal errors have also been estimated with the values derived from multiple spectra of the same stars. A forthcoming paper (Y. Pan et al., 2018, in prep.) will focus on the same topic {with a database} more than twice the previous spectral library, from 61\,226 to 173\,971. The number of common stars which have been observed by both {\sl Kepler} and LAMOST is { 76\,283}, a number { larger than} twice the data size of \citet{2016ApJS..225...28R}. In addition, the previous spectra have been {re-}analyzed with the updated version of the LASP code, v2.9.7, where the determination of uncertainty of radial velocity is improved significantly. The new calibration examines the reliability of the previous results provided by \citet{2016ApJS..225...28R}, to establish more generic calibration formulae.

{Other groups, including the ``European''  and ``American'' teams, are applying their own codes to these spectra for the determination of basic stellar parameters.} The ``European team'' (Molenda-\.Zakowicz et al., in prep) is determining the atmospheric parameters, the spectral classification, the projected rotational velocity (for ultrafast rotators only) and activity indicators with the latest adapted version of the code ROTFIT, for the entire database extending the results of their previous work \citep{2016A&A...594A..39F}. The ``American team'' (R. O. Gray et al., 2018, in prep.) is 
performing an accurate MK classification with the automatic code MKCLASS, which is independent of the determination of atmospheric  parameters \citep[see details in][]{2016AJ....151...13G}.

{The current LASP and the above pipelines have not yet incoporated the templates} 
for hot and highly evolved stars, including the O and B type main-sequence stars, the white dwarfs and the hot subdwaf stars. With the previous LAMOST data (including the LK-project's), several independent works have been concentrated on highly evolved compact stars, aiming at the characterization and determination of stellar parameters for white dwarfs \citep{2015MNRAS.454.2787G,2013AJ....145..169Z} as well as hot subdwafs \citep{2016ApJ...818..202L}. 
{These stars, which are not analyzed by the LASP code,
are scientifically valuable when their parameters are derived by
specialized outside pipelines or models} \citep[see, e.g.,][]{2017ApJ...847...34S}.

In line with the initial goals for {the} LK-project, there are fruitful results {being generated based on this} catalog together with the {\sl Kepler} photometry. {While i}n depth studying the {\sl Kepler} targets based on the whole database, a cross-identification was established between the catalog of the LAMOST DR4/DR5 and the {\sl Kepler} archive, with a tolerance of three arcsec in coordinates separation (A. Luo et al., 2018 in prep.), which is useful to give the contamination factor of the targets that are affected by their neighbor sources during the {\sl Kepler} observations {(see also Table~5 of Paper\,\RNum{1} which lists this value for the DR3 data)}. For the stars with large contamination factor, one should be extremely careful since those targets may be very close to their neigbor objects and a wrong star observed by LAMOST cannot be ruled out.

{The astronomical community can exploit the spectra of such a vast database for researches in many different fields}, such as stellar activity, binaries and asteroseismology. \citet{2017ApJ...849...36Y} provided a comprehensive investigation of stellar flare events in M-dwarfs in the {\sl Kepler} field with the H$\alpha$ emission lines illustrated from the LAMOST spectra. There are 483 stars with multiple (5+) spectr{a} which can be used to check the variations of radial velocities, serving as an independent technique to discover new binary systems and to solve their orbital parameters \citep[see, e.g.,][]{2018MNRAS.477.2020C}. A major goal of the LK-project is to provide accurate parameters for pulsating stars since their pulsations can be measured to a precision of a few tenth of {a} nano hertz from the {\sl Kepler} photometry \citep[see, e.g.,][]{2016A&A...585A..22Z}, which are key input parameters for the seismic modeling of pulsating stars. 

A drawback of the LK-project is that the {\sl Kepler} orignal field can only be reached by LAMOST during the summer season when the weather conditions typically show a high precipitation due to the Monsoon (Paper\,\RNum{1}). Therefore, to fully extend the capacity of LAMOST, providing spectra for targets with high-quality photometry, we select six {\sl K2} campaigns \citep[i.e., C0, C1, C4, C5, C6 and C8;][]{2014PASP..126..398H} whose declinations are higher than $-10^{\circ}$, hence reachable by LAMOST to carry out the LAMOST-{\sl K2} project observations. Each field is divided into 14 subfields, similar to the LK subfields. During a period from 2014 to 2017, a total of 151 plates had been observed over 108 nights, with a collection of 291\,956 spectra for 222\,926 stars (J. Wang et al., 2018, in prep.). The calibration of $\log g$ between the asteroseismic determination and the LASP code have been obtained by {\color{blue} Zhang et al. (2018, in prep.)}. We also mention that a current proposal of LAMOST has been approved for the follow-up observations to the region near the north ecliptic pole where the {\sl TESS} mission will observe continuously with a duration {of} about one year \citep{2014SPIE.9143E..20R}.  
 
As LAMOST had finished its first five-year regular survey in 2017 summer, several insteresting proposals had been  approved, in particular observations relat{ed} to {the} LK-project. One of them is the time-series spectroscopic observations for the targets in the field of {\sl Kepler} and {\sl K2}, with 16 middle-resolution ($R\sim7500$) spectrographs but for two different arms, with the wavelength ranges of 630---680\,nm and 495---535\,nm, respectively. {Observations of the test plate (i.e., LK07) have been completed on 2018 May 24, 28-31. The preliminary data set of parameters contains about 1800 targets with most of them revisited of 30 times. A forthcoming paper will concentrate on the test results. All the data will be available publicly after 2020 September along with the DR6.}

\section{Summary}
The LK-project, as a support for follow-up spectroscopic observations for {\sl Kepler} photometry, aims at providing spectra for investigating a variety of physics of the stars observed by {\sl Kepler}. During the first five-year regular survey phase, the original {\sl Kepler} field has been observed by LAMOST on 83 plates over 49 nights from 2012 June to 2017 June, accumulating 227\,870 spectra of 156\,390 stars. 
{ From the cross identification of $\sim200\,000$ targets with available {\sl Kepler} photometry,}
we find { 76\,283} stars in common between LAMOST and {\sl Kepler} with constraints on the coordinate separation { ($d < 3.0$}~arcsec). 

At the current stage, we have obtained stellar parameters with 173\,971 spectra of 126\,172 stars of type A, F, G and K star with the LASP pipeline v2.9.7, which provide hundreds of candidates of high-velocity stars and metal-poor or very metal-poor stars. {A process of homogenization between our data and the products of other works is ongoing. This work will certainly provide us with a} more robust estimation of the derived parameters. These spectra will be useful for the entire astronomical community in particular for investigating planetary science and stellar physics of {\sl Kepler} targets.

Although LAMOST has completed its first five-year regular survey, the LK-project will be continued in the next years to provide more spectra. In parallel to the LK-project, there are several approved programs to use LAMOST for follow-up observations of targets of already operating and forthcoming space missions such as {\sl K2} and {\sl TESS} satellites. In order to provide time-series spectroscopy, one footprint on {\sl K2} C0 field has been continuously observed for more than ten times. A very similar project is under test while  LAMOST is equipped with intermediate-resolution spectrographs. Astronomers who are interested in those spectra can contact the members of the LAMOST consortium\footnote{Send requests to Dr. Ali Luo (lal@bao.ac.cn)}.

\acknowledgments
The Guoshoujing Telescope (the Large Sky Area Multi-object Fiber Spectroscopic Telescope LAMOST) is a National Major Scientific Project built by the Chinese Academy of Sciences. Funding for the project has been provided by the National Development and Reform Commission. LAMOST is operated and managed by the National Astronomical Observatories, Chinese Academy of Sciences. WKZ hosts the LAMOST fellowship as a Youth Researcher which is supported by the Special Funding for Advanced Users, budgeted and administrated by the Center for Astronomical Mega-Science, Chinese Academy of Sciences (CAMS). WKZ, JNF, YP, RYZ, JTW and MQJ acknowledge the support from the National Natural Science Foundation of China (NSFC) through the grant 11673003, {11833002} and the National Basic Research Program of China (973 Program 2014CB845700). JM-\.Z acknowledges the Polish National Science Centre grant no. 2014/13/B/ST9/00902 and the Wroclaw Centre for Networking and Supercomputing grant no.224. {S.D. acknowledges Project 11573003 supported by NSFC and the LAMOST Fellowship, which is supported by Special Funding for Advanced Users, budgeted and administrated by CAMS. The work presented in this paper is supported by the project "LAMOST Observations in the {\it Kepler} field" (LOK), approved by the Belgian Federal Science Policy Office (BELSPO, Govt. of Belgium; BL/33/FWI20).}

\software{LASP \citep[v2.9.7;][]{2011A&A...525A..71W,2015RAA....15.1095L}}

\bibliographystyle{aasjournal}

\end{document}